# Dataset of Tensile Properties for Sub-sized Specimens of Nuclear Structural Materials

***Authors***

Longze Li[1], John W. Merickel[2], Yalei Tang[2], Rongjie Song[2], Joshua E. Rittenhouse[2], Aleksandar Vakanski[1], Fei Xu[2]

**Affiliations**

[1] Department of Computer Science, University of Idaho, Moscow, ID, United States

[2] Idaho National Laboratory, Idaho Falls, ID, United States

Corresponding authors: Aleksandar Vakanski (vakanski@uidaho.edu), Fei Xu (Fei.Xu@inl.gov)

***Abstract***

Mechanical testing with sub-sized specimens plays an important role in the nuclear industry, facilitating tests in confined experimental spaces with lower irradiation levels and accelerating the qualification of new materials. The reduced size of specimens results in different material behavior at the microscale, mesoscale, and macroscale, in comparison to standard-sized specimens, which is referred to as the "specimen size effect." Although analytical models have been proposed to correlate the properties of sub-sized specimens to standard-sized specimens, these models lack broad applicability across different materials and testing conditions. The objective of this study is to create the first large public dataset of tensile properties for sub-sized specimens used in nuclear structural materials. We performed an extensive literature review of relevant publications and extracted over 1,000 tensile testing records comprising 55 parameters including material type and composition, manufacturing information, irradiation conditions, specimen dimensions, and tensile properties. The dataset can serve as a valuable resource to investigate the specimen size effect and develop computational methods to correlate the tensile properties of sub-sized specimens.

## Background & Summary

The development of advanced nuclear reactor concepts aims to enhance the safety, economics, waste management, and non-proliferation security of commercial nuclear power reactors [1]. The advanced reactor designs typically operate under more extreme conditions with higher temperatures and increased radiation levels compared to current light water reactors [2][3]. Such conditions necessitate the timely development of suitable advanced materials and processes. Unlike materials development in other sectors which typically spans a few years [4], the development and qualification of nuclear materials are prolonged due to the need for extensive testing and experiments to assess in-reactor performance. Accelerated irradiation testing thus is crucial for rapidly achieving high irradiation doses and understanding the behavior and irradiation responses of new materials [5].

Small-scale mechanical testing with sub-sized specimens in the nuclear industry enables faster irradiation tests and reduces the time for deployment of new materials [5]–[7]. Importantly, small-scale mechanical testing allows for greater utility of constrained experimental spaces within nuclear test facilities, and enables researchers to conduct a greater number of experiments while simultaneously reducing the radiological dose emitted by irradiated materials, which is directly proportional to their volume [8][9].

Despite the advantages of mechanical testing with sub-sized specimens, this approach is marked with several challenges, such as the demand for enhanced testing precision and advanced measurement methods, sensitivity to testing conditions and dimensional deviations, as well as increased impact of the sample preparation, processing techniques, and surface defects on the test results. Consequently, mechanical properties of sub-sized specimen materials are often reported to exhibit greater variability compared to standard-sized specimens [9]–[11]. The varying material behavior due to the reduced dimensions of sub-size specimens in comparison to standard-sized specimens is commonly referred to as the specimen size effect or scaling effect [5][6][12]–[14]. In general, this effect encompasses a broad range of other factors impacting measured material properties beyond specimen size and geometry, and involves microstructural changes and inhomogeneity, surface effects, and anisotropy resulting from the differences in the microstructure and crystallographic texture [10]. Failing to account for the specimen size effect can lead to incorrect estimates of material properties and load-bearing capabilities, which can potentially result in catastrophic consequences in critical applications, such as the nuclear industry.

Extensive research has been conducted to develop models that link the test results from sub-sized specimens to those of standard-sized specimens [5]–[15]. Techniques for conversion of elongation based on the Bertella-Oliver formula [14] or Barba's law [15] have been broadly adopted for correcting the measurements of small-size specimens, provided that the bulk material properties are maintained. The inverse Finite Element Method (FEM) has also been utilized to predict true stress-strain relationships from engineering strain-stress curves based on minimizing the deviations in contact-type strain measurements [16][17]. The FEM method is also used for predicting the material behavior beyond the yield point and after necking [18]. Additionally, several application standards provide recommendations on conversion methods for the mechanical properties of sub-sized specimens and standard-sized specimens [19]. On the other hand, existing analytical models and conversion methods are typically developed for specific materials and tests and/or introduce certain assumptions that limit their applicability across different testing conditions and materials. Consequently, these models may produce inaccurate results under varying conditions and for different materials [20]. Furthermore, many analytical models do not provide means to evaluate the importance of test conditions or individual input features on the specimen size effect.

Recent progress in Machine Learning (ML) offers unique potential to advance materials science [21]–[23], due to the capacity for modelling complex combinatorial spaces of multi-physics and multi-scale mechanisms, which traditional analytical methods are unable to resolve or can only address at a significant computational expense. These ML advantages can be effective in addressing the challenges related to specimen size effect in mechanical tests [24][25]. Despite the promising capabilities, a major obstacle that hinders the wide application of ML-based approaches in materials science is the lack of large, curated datasets with well-organized and consolidated information for various materials and testing conditions. The scattered nature of helpful information in the published literature on mechanical tests and other material information prevents the direct application of data-driven ML methods.

The objective of this study is to systematically gather reliable data for investigating the specimen size effect on the tensile properties of nuclear structural materials. The focus is on studying the influence of specimen dimensions and geometry on mechanical properties such as yield strength, ultimate tensile strength, uniform elongation, and total elongation. The dataset was created through an extensive literature review of scientific articles and databases. The search inclusion criteria targeted peer-reviewed studies on tensile testing of sub-sized specimens, providing quantitative data on tensile properties relative to specimen size. The extracted data points from the literature review were organized into a tabular format database containing 1,050 tensile testing records with 55 parameters, including material type and composition, manufacturing information, irradiation conditions, specimen size and dimensions, and tensile properties. Materials science experts conducted systematic checks to validate the collected data, ensuring accuracy in the material type, manufacturing processes and treatment methods, and testing conditions, as well as verifying the chemical composition and other pertinent information. Our team performed statistical analyses to identify and address data outliers, ensuring the reliability of the dataset.

The contributions of this work are as follows:

- Presents the first large curated public dataset of tensile properties for sub-sized specimens of three common groups of nuclear structural materials.
- Involves a comprehensive literature review to collect over 1,000 data records of tensile tests encompassing 55 parameters, related to material type and composition, manufacturing information, irradiation conditions, specimen size and dimensions, and tensile properties.
- Provides a valuable resource readily available to researchers and engineers for developing computational and ML-based methods for establishing correlations between the tensile properties of sub-sized and standard-sized specimens.

## Methods

Sub-sized specimens are also referred to as small-size, small-scale, mini-size, or miniaturized specimens in various studies [7][26]. Sub-sized specimens were initially developed in the 1970s for testing materials for nuclear reactors due to limited space for experiments in neutron facilities [27][28]. In addition, smaller specimen volumes reduce induced activation, thereby reducing environmental pollution. There has been an increasing demand for using sub-sized specimens in other domains and industries where a limited amount of material is available, such as for assessing the residual life of in-service components by scooping out a small volume of material. Other related applications include evaluating the mechanical properties of micro-electromechanical systems, additively manufactured parts, in rapid alloy prototyping, for measuring local properties in welded parts, and other similar uses [29]–[32].

Miniaturized Tensile Test (MTT) is a technique for evaluating the tensile properties of materials using sub-sized specimens with the smallest dimension ranging from hundreds of microns to several millimeters [7][33]. MTT is a result of a sustained research and engineering effort based on international collaborations toward standardizing tensile testing with sub-sized specimens. MTT faces various challenges that stem from the specimen size effect, requirements for specialized high-precision equipment and testing procedures, and increased sensitivity to dimensional deviations, specimen fabrication and preparation, surface defects, and other factors. Although numerous studies have been dedicated to addressing the challenges and improving the reliability of MTT, it remains an open topic of research. The following sections present a brief overview of tensile testing and discuss the related challenges of the specimen size effect and design of sub-sized specimens for MTT.

### *Tensile Test Basics*

Tensile test is a mechanical test in which a specimen of material undergoes controlled tension until it fractures. The specimen is typically shaped as a cylindrical profile with a uniform cross-sectional area in the middle (also known as dog-bone shape) to ensure uniform distribution of force and predictable failure locations. The specimen is placed in a tensile testing machine that clamps each end, and longitudinal pulling force is applied stretching the specimen at a consistent rate until it fractures [15].

The main outcome of a tensile test is the engineering stress-strain curve shown in Figure 1, which illustrates the material behavior as the load is applied. The stress value $\sigma$ is derived by dividing the applied axial force $F$ with the original cross-sectional area $A_0$ of the specimen.

$$\sigma = \frac{F}{A_0} \tag{1}$$

The engineering strain $e$ is unitless, and it is computed using equation (2), where $L_i$ represents the instantaneous length and $L_0$ is the initial length of the specimen before the test starts.

$$e = \frac{L_i - L_0}{L_0} \tag{2}$$

The engineering stress-strain curve provides important information about the mechanical properties of a material. *Yield strength* (YS) $\sigma_s$ is defined as the maximum stress that a material can withstand before it begins to deform plastically. When the applied stress is less than $\sigma_s$, the material will recover its original shape once the applied force is removed. *Ultimate tensile strength* (UTS) $\sigma_u$, also referred to as tensile strength, is the highest stress that a material can endure while being stretched until it breaks, and it corresponds to the peak stress on the stress-strain curve just before the material fails and fractures. Once the material reaches UTS, it has achieved the peak capacity to handle stress and starts to neck, where necking refers to the drastic local plastic deformation. *Uniform elongation* (UE) $e_u$ is the maximum elongation the material can achieve before necking happens. *Total elongation* (TE) $e_f$ is also known as fracture elongation or total elongation at fracture, and it represents the sum of the uniform elongation $e_u$ and *post-necking elongation* (PE) $e_p$.

$$e_f = e_u + e_p \tag{3}$$

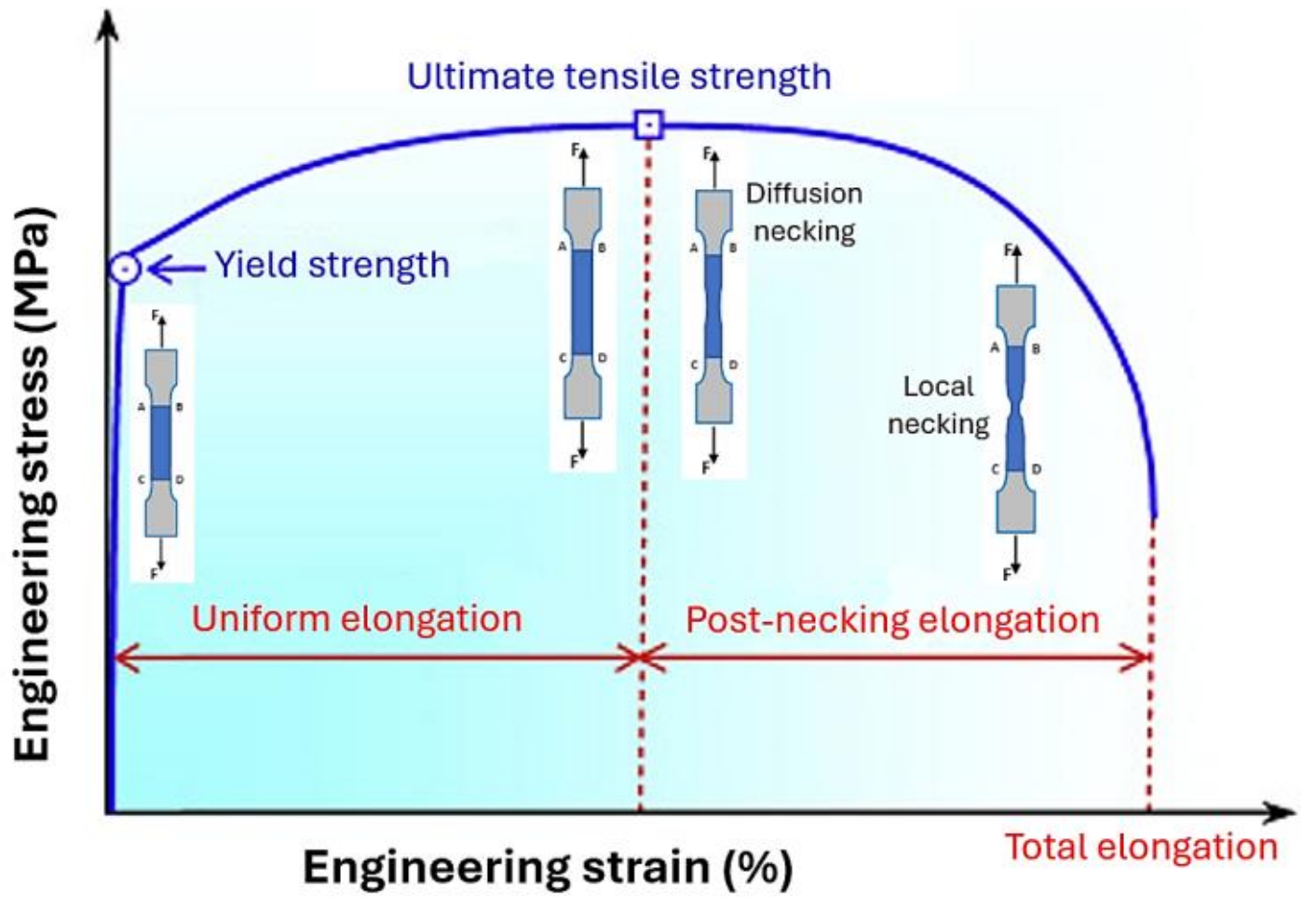

**Fig. 1.** Engineering stress-strain curve in tensile test.

While engineering stress-strain curves are generally used in assessing the tensile mechanical properties of materials, true stress-strain curves more accurately reflect the constitutive behavior of materials and are typically used for theoretical analysis, high precision applications, and detailed mechanical calculations. The true stress $\sigma_t$ is calculated as the applied load $F$ divided by the instantaneous cross-sectional area $A_i$ of the sample.

$$\sigma_t = \frac{F}{A_i} \tag{4}$$

The true strain $e_t$ is calculated as the natural logarithm of the instantaneous length $L_i$ divided by the original length $L_0$.

$$e_t = \ln\left(\frac{L_i}{L_0}\right) \tag{5}$$

### *Specimen Size Effect*

As stated before, the *specimen size effect* refers to the difference in the measured mechanical properties for specimens with reduced dimensions in comparison to standard-sized specimens. As the specimen size is reduced beyond a certain threshold, correlating the properties of sub-sized specimens with those of standard-sized specimens using analytical methods becomes challenging. The challenges arise not only from the inconsistencies in scaling the material microstructure relative to the changes in material properties at the mesoscale and macroscale, but also from the differences in the specimen geometry at the microscale due to the impact of residual stresses and surface roughness.

Two main design types of sub-sized specimens include flat (rectangular) specimens and round (cylindrical) specimens, as depicted in Figure 2.

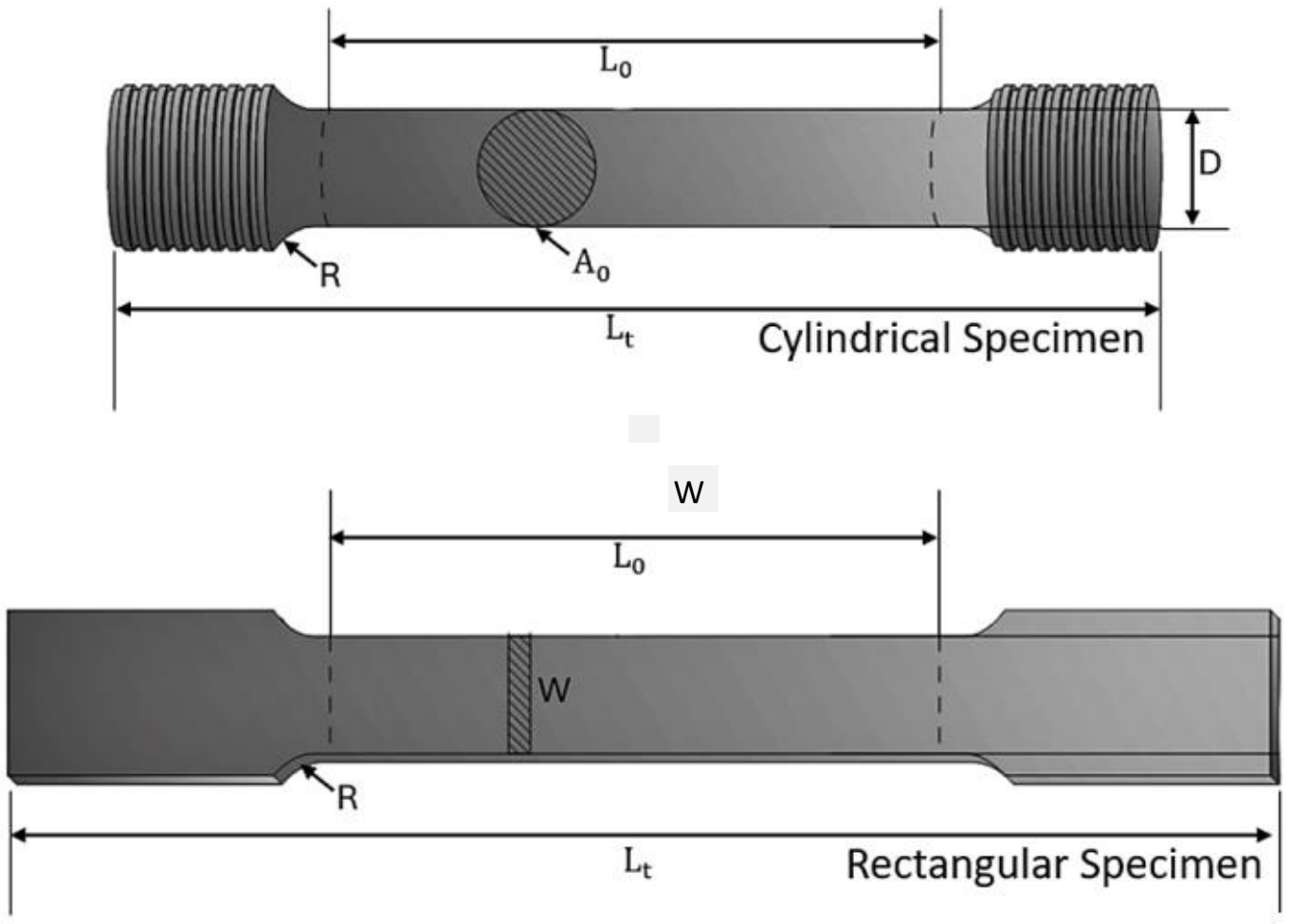

**Fig. 2.** Tensile specimen examples with cylindrical (round) and rectangular (flat) shape. $L_0$ is gauge length, $L_t$ is total length, and R is the radius of the fillet. $A_0$ denotes the cross-sectional area for the cylindrical specimen and W denotes the width for the flat specimen.

In flat sub-sized specimens, strength and ductility are affected when the specimen thickness is reduced below a certain value. Specifically, the thickness impacts yield strength, ultimate tensile strength, uniform elongation, post-necking elongation, and total elongation. The critical thickness largely depends on the thickness-to-grain size ratio $T/GS$, and for most materials the critical value of the ratio is between 6 and 10 [34][35]. Particularly, when the $T/GS$ is below the critical value, the increase in surface layer grains causes increased dislocation density that cannot maintain the strain hardening rate of the bulk material, and results in decreased flow stress and uniform elongation. When the $T/GS$ ratio is above the critical value, strength and ductility remain nearly constant.

Specimen width also influences the test results for ultimate tensile strength, post-necking elongation, and total elongation in flat sub-sized specimens, even when the thickness is sufficient to represent bulk material properties. A critical width value is typically determined based on the width-to-thickness ratio $W/T$ being approximately 5. When $W/T$ is above 5, ultimate tensile strength decreases with increasing $W/T$, although yield stress remains unaffected by changes in specimen width [7][36]. Decreasing $W/T$ leads to increased post-necking elongation in structural steels due to variations in the necking behavior that changes from localized to diffuse necking, resulting in increased overall total elongation [37]–[40].

Specimen gauge length significantly impacts ductility measurements, where shorter gauge lengths lead to increases in post-necking elongation and total elongation [41]–[43]. A critical value based on the ratio of the original gauge length to the square root of the cross-sectional area $L_0/\sqrt{A}$ of 5.65 is typically adopted [7]. The amount of elongation that occurs during post-necking deformation is not influenced by gauge length, therefore uniform elongation is unaffected by gauge length.

The impact of specimen diameter and gauge length on material properties for round sub-sized specimens is similar to the impact of thickness and gauge length for flat sub-sized specimens [44]. That is, the diameter must contain a critical number of grains to accurately represent bulk properties, where critical values of the diameter-to-grain size ratio $D/GS$ from 6 to 10 are typically adopted. Similarly to flat specimens, gauge length affects the ductility of the material. For instance, Yuan et al. [45] reported that reductions in the gauge length-to-diameter ratio $L_0/D$ in pure copper increased post-necking and total elongation, while uniform elongation remained consistent regardless of specimen size.

The critical values for the specimen size effect based on the important ratios of sub-sized specimens and the impacted tensile properties are summarized in Table 1.

**Table 1**. Critical values for sub-sized specimen dimensions and impacted tensile properties.

| Specimen Geometry Type | Specimen Dimensions Ratio | Critical Value | Impacted Tensile Properties |
|---|---|---|---|
| Flat | Thickness-to-Grain Size ($T/GS$) | $6-10$ | YS, UTS, UE, PE, TE |
| Round | Diameter-to-Grain Size ($D/GS$) | | |
| Flat | Width-to-Thickness ($W/T$) | 5 | UTS, PE, TE |
| Flat | Length-to-Square Root of Area ($L_0/\sqrt{A}$) | 5.65 | PE, TE |
| Round | | | |

## *Sub-sized Specimen Design*

Unlike the design of conventional standard-size specimens that are based on standards by ASTM (American Society for Testing and Materials), ISO (International Standards Organization), JIS (Japanese Industrial Standards), and related standardization organizations which provide strict specifications for the specimen dimensions and aspect ratios, currently standards for the design of sub-sized specimens are not available. The need for standardized design methodology has been widely acknowledged, and there are undergoing efforts in standard development by several organizations. Also, multiple MTT studies provide guidelines and recommendations for the design of sub-sized specimens [5].

MTT guidelines for the design of sub-sized specimens are summarized as follows [7][26][46][47]. (a) The specimen should at least have 6 to 10 grains in either thickness or diameter to represent the bulk material properties. (b) The specimen geometry should satisfy requirements for the aspect ratios $W/T$ and $L_0/\sqrt{A}$ in order for the test results to be consistent with the standard-size specimens. (c) The design of the specimen should aim to reduce data variability, as sub-sized specimens often exhibit greater variability in test results compared to standard-sized specimens. (d) The volume of the specimens should be kept to a minimum to decrease the radioactivity of irradiated materials and mitigate risks during sample collection. (e) Specimen manufacturing costs should be considered in the design, e.g., flat sub-sized specimens are preferred because they are easier to manufacture compared to round specimens. (f) Testing and measurement equipment should be considered to ensure that the specimen design is a suitable fit for the available equipment.

Many different designs of sub-sized specimens presently exist, due to the lack of standardization. In general, the specimen design is typically either based on direct scaling of the dimensions of standard-sized specimens, or indirect scaling by applying additional adjustments to certain dimensions. The above requirements for lower bounding the minimal dimensions and other guidelines need to be observed in both cases. The former approach results in sub-sized specimens that have proportional geometry to dog-bone shaped standard-size specimens. The tensile properties of such specimens are often reported to be consistent with standard-sized specimens, although the size effect requires cautiousness and validation. In addition, proportional specimens may still have relatively large gauge lengths in comparison to the other dimensions. Further reduction in the specimen volume resulting in lower irradiation dose in the nuclear industry can be achieved by the latter approach, in which sub-sized specimens are designed with geometry that is not proportional to the standard-size specimens. Data analysis methods based on elongation conversion methods or inverse FEM have been used by researchers in designing the specimen geometry [48]. In most cases, these specimen designs include reduced gauge length, and are hence more suitable for tests in constrained irradiation facilities. Due to the specimen size effect, such design requires a more rigorous validation step, that is typically material dependent.

Examples of sub-sized specimens from the SS and SS-J series are shown in Figure 3 [49]. SS-1 specimen design was developed for the material science irradiation program in the Experimental Breeder Reactor-II reactor in the 1980s [50]. SS-2 design with smaller dimensions was subsequently developed to reduce the volume for irradiation tests [51]. SS-3 and SS-J3 designs with reduced gauge lengths of 7.62mm and 5 mm, respectively, were developed in the 1990s for irradiation in the fast flux test facility [49]. Sub-sized specimens with gauge lengths as low as 2 mm have also been proposed for fitting into constrained irradiation volume and with reduced radiation dose [52].

Besides the dog-bone shape sub-sized specimens, other proposed shapes include bow-tie shape [53] and dumb-bell shape [54]. These specimens can address some of the challenges in tensile testing of miniature specimens, such as gripping of the specimen. On the other hand, these designs require specialized testing equipment and processing techniques, there is a small amount of tensile data available, and demand careful validation across different materials. Due to these reasons, dog-bone shape sub-sized specimens are still the norm in tensile testing.

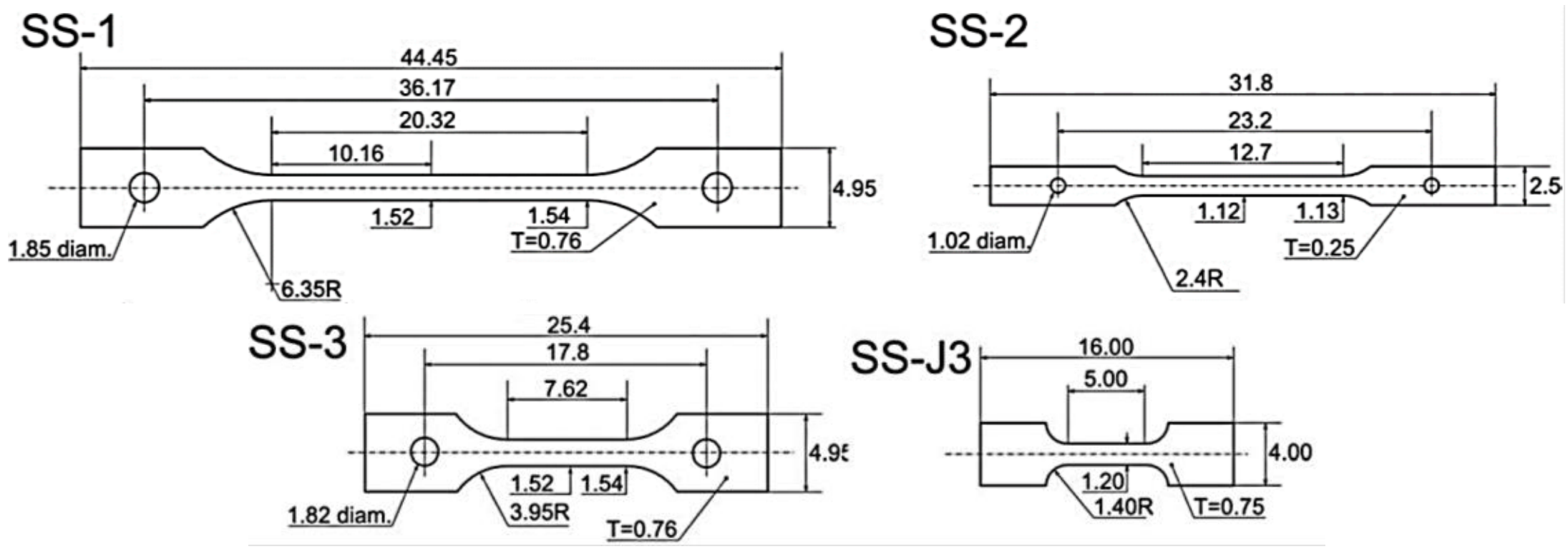

**Fig. 3**. Geometry and dimensions of sub-sized specimens from the SS and SS-J series. All dimensions are in millimeters. T denotes thickness. (Reprinted, with permission from [49]. Copyright ASTM International, http://www.astm.org).

# Data Records

## *Data Acquisition*

The data was collected through a comprehensive search process of relevant scientific articles in the published literature. The search strategy was based on using various combinations of the following query terms: “tensile test,” “sub-sized specimen,” “mini specimen,” “miniature specimen,” “small specimen,” “size effect,” “stainless steel 316,” “SS316,” “reactor pressure vessel steel,” “zirconium alloy,” and “zircaloy.” The quotation marks for the above terms indicate that the search queries comprised more than a single individual keyword. We used the Google Scholar website search engine, Google Chrome web browser search engine, and Microsoft Edge web browser search engine. Based on an extensive literature search using the listed specific terms, we identified 72 scientific peer-reviewed articles in total containing potentially relevant data for the task at hand.

In the next step, we applied three inclusion criteria to the retrieved articles for further refining the search process and identifying articles that contain relevant data. Inclusion criteria involved: (a) the article provides quantitative results of tensile tests with sub-sized specimens regarding yield strength, ultimate tensile strength, ultimate elongation, and total elongation; (b) the article provides information regarding the dimensions and geometry of used sub-sized specimens; and (c) the specimen material belongs to the following groups of nuclear structural materials: stainless steel (SS) 316 (grades SS316, SS316L, SS316LN, and SS316H), reactor pressure vessel steels (such as SA-508, SA-533, 20MnMoNi55), and fuel cladding Zirconium alloys (Zircaloy-2, Zircaloy-4, and Zr2.5Nb). Articles lacking detailed experimental tensile test data, not employing sub-sized specimens, or focusing on different materials were excluded from the data extraction. This step resulted in 24 identified articles, listed in Table 2.

For the articles that provide structured tabular data of tensile test results and information related to specimen size, manufacturing treatment, and other relevant data in the form of tables, spreadsheets, or databases, we manually extracted the data and stored them in an Excel workbook. For the articles where the data appear solely in graphical form, we employed the plot digitization software WebPlotDigitizer [55] to manually extract the data for each point in the graphs. Examples include graphs of the distribution of the tensile properties as a function of the specimen dimensions. Material science experts double-checked the extracted tabular and graphical data to validate the accuracy and reduce the likelihood of human error. The total number of extracted tensile test records is 1,050. The number of records, material type, and the articles from which they are extracted are presented in Table 2.

**Table 2**. Acquired number of data records and material type in the dataset per reference.

| Number of Records | Material Tested | Data Source | Dataset |
|---|---|---|---|
| 17 | SS304L, FeCrAl, A-718, Al-6061 | Gussev et al. (2017) [5] | Byun et al. (1998) [56] |
| 4 | Zircaloy-4 | Pierron et al. (2003) [57] | Experimental data |
| 36 | SS316L, F82H, SA-533 | Gussev et al. (2015) [49] | Experimental data |
| 421 | JPCA, JFMS | Kohno et al. (2000) [36] | Experimental data |
| 24 | SS304, S235, S355 | Rund et al. (2015) [33] | Experimental data |
| 11 | SA-508 | Byun et al. (1998) [56] | Experimental data |
| 2 | Ultra-fine grain low carbon steel | Alsabbagh et al. (2013) [58] | Experimental data |
| 14 | Experimental low-carbon steel, Ti6Al4V, X14CrMoVNbN10, P91, EN AW, 6005 T6 Copper 99.99 % | Džugan et al. (2015) [59] | Experimental data |
| 8 | Ti6Al4V | Van Zyl et al. (2016) [60] | Experimental data |
| 25 | DP600, DP800, SS316L | Zhang et al. (2021) [13] | Experimental data |
| 26 | 20MnMoNi55, CrMoV, SS304LN | Kumar et al. (2014) [61] | Experimental data |
| 22 | Zircaloy-2, Zr-2.5Nb | Balakrishnan et al. (2014) [62] | Experimental data |
| 9 | DC01 | Konopík et al. (2018) [63] | Experimental data |
| 3 | CP AL (AL-99.6 wt.%) | Lanjewar et al. (2019) [64] | Experimental data |
| 4 | UNS S31035, Sanicro 25 | Dymáček et al. (2018) [54] | Experimental data |
| 3 | Ti6Al4V | Sikan et al. (2023) [65] | Experimental data |
| 72 | SS316 | Klueh (1985) [50] | Experimental data |
| 40 | SS304, SS316 | Igata et al. (1986) [66] | Kestenbach et al. (1976) [67] |
| 5 | SS316L | Roach et al. (2020) [68] | Experimental data |
| 238 | SS316 | Miyahara et al. (1985) [69] | Experimental data |
| 22 | SA-508 | Yin et al. (2023) [70] | Experimental data |
| 4 | SS316 | Do Kweon et al. (2021) [71] | Experimental data |
| 4 | SS316 | Mishra et al. (2021) [72] | [73]-[84] |
| 24 | SS304, SS304L, SS316L | Seo et al. (2022) [85] | Experimental data |

### *Data Post-Processing*

We applied several post-processing steps to ensure that all records in the dataset provide consistent information. Specifically, the extracted data from various articles contained information reported in different measurement units, such as testing or treatment temperatures reported in Celsius, Fahrenheit, and Kelvin degrees. We applied unit conversion for several features in the collected dataset to ensure consistency in the data records. Other similar examples include converting the strain-rate information for the tensile test from strain/minutes to strain/seconds, as well as cases where the articles only disclosed the used machine speed and we derived the strain rate based on the sample length and machine speed. Furthermore, the information for the material composition in several papers included the term "balance" to indicate that the remainder of the composition is made up of the specified primary element providing the base structure after accounting for all other elements (e.g., Iron (Fe) in SS316). We calculated the composition of the primary element to ensure that the weight percentage of all elements in the provided composition data sums to 100%, and we replaced the term "balance" with the calculated value. We confirmed the accuracy of the composition for each record in the dataset by ensuring that the sum of the elements is consistently 100% with no outliers.

### *Data Organization*

The extracted data are stored in a Microsoft Excel workbook format, archived in the is available at the Materials Cloud Archive [86], as well as in the data repository on GitHub (link: https://github.com/avakanski/Subsized-Specimens-Tensile-Properties). The structure of the dataset is schematically depicted in Figure 4. Each row corresponds to retrieved information for a single tensile test. The information for each record consists of 55 features (columns), divided into the following categories: Reference, Material Type and Composition, Manufacturing and Treatment Information, Irradiation Conditions, Specimen Dimensions, Tensile Test Conditions, and Tensile Properties. The columns in the Reference category list the article from which the data were extracted. The Material Type and Composition columns record the material type of the sub-specimens and their chemical composition listing the elements given in weight percentage. The category Manufacturing and Treatment Information documents the type of treatment, post-treatment temperature (°C) and time (hours), preparation (finishing processes), microstructure, morphology, and grain size (μm), and similar information. The Irradiation Conditions columns, if applicable to the tensile test record, provide information regarding the irradiation dose (milli-displacement per atom), temperature (°C), and time (hours). The information for all columns in the Manufacturing and Treatment Information and Irradiation Conditions were not provided in all articles, and as a result, some test records have missing information (e.g., for microstructure, or grain size). Specimen Dimensions are provided in millimeters and present information regarding the gauge length, width, thickness, fillet radius, and aspect ratios of the dimensions of sub-sized specimens. Tensile Test Conditions provide information regarding the temperature of the tensile test (°C), and the strain rate ($1/s$) of the test. Finally, Tensile Properties include data on yield strength (MPa), ultimate tensile strength (MPa), uniform elongation (%), and total elongation (%). Each category label is color-coded in the Excel workbook.

Besides the main worksheet containing the data, the Excel workbook contains a worksheet References that lists the papers from which the data was extracted, and a worksheet Column Descriptions that provides descriptions of the columns in the dataset.

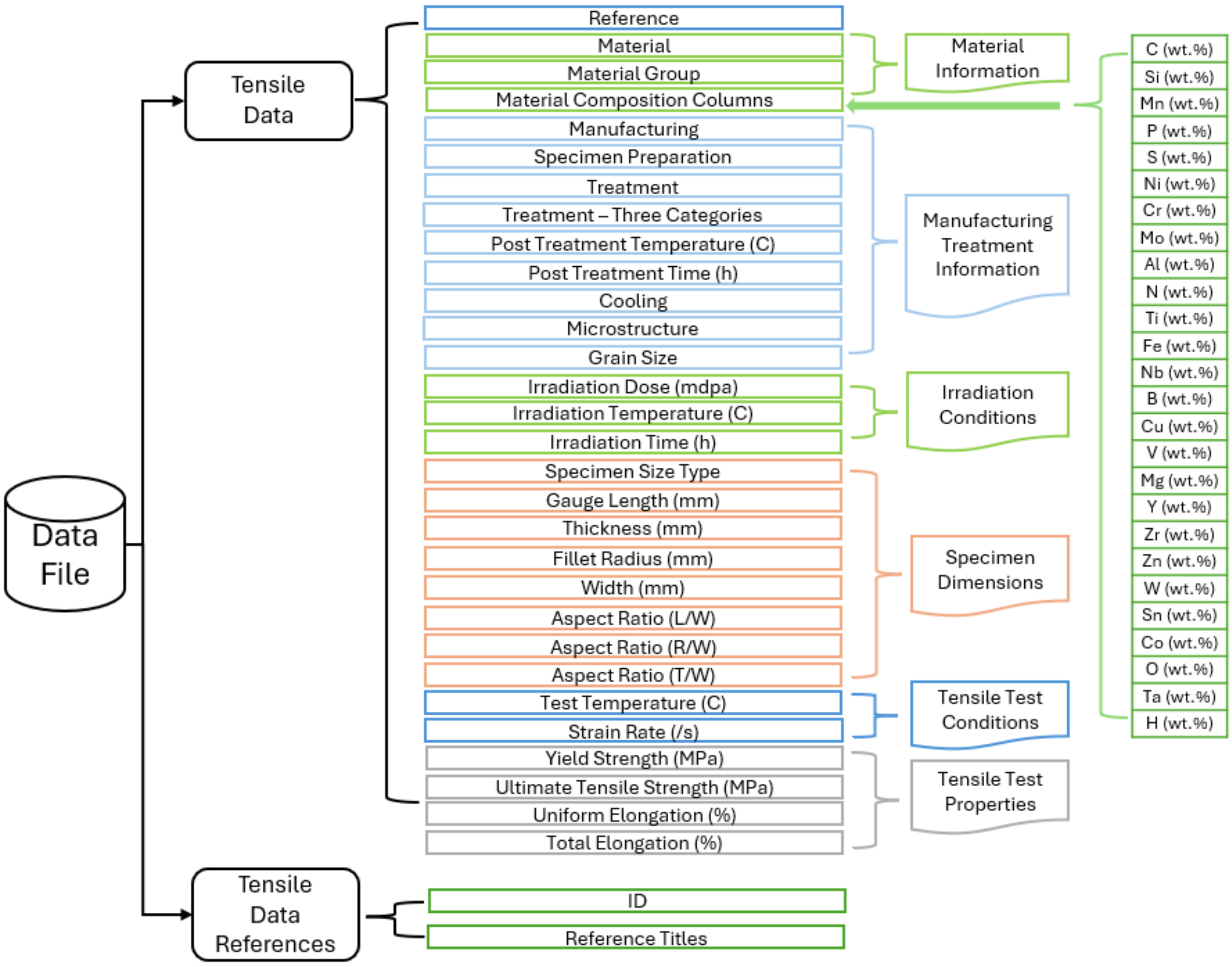

**Fig. 4**. Organization of the dataset by columns and data categories.

## Specimen Size

The dimensional information of the sub-sized specimens in the dataset, related to the gauge length, width or diameter, and thickness or diameter, along with the specimen geometry and number of records in the dataset is presented in Table 3. Histograms of the distribution of gauge length, width, and thickness for the specimens in the dataset are shown in Figure 5.

**Table 3**. Dimensions of sub-sized specimens in the dataset. Note: the brackets in some of the fields indicate that multiple values are used for that specimen dimension in the corresponding reference.

| Sub-sized Specimen Type | Gauge Length (mm) | Width (mm) or Diameter (mm) | Thickness (mm) or Diameter (mm) | Specimen Geometry | Number of Records |
|---|---|---|---|---|---|
| SS Mini 1S1, SS Mini 1E1 [5] | 2.55 | 0.8 | 0.4 | Flat | 2 |
| SS Mini 1S2, SS Mini 1E2 [5] | 2.55 | 0.8 | 0.6 | Flat | 2 |
| SS Mini 2S1, SS Mini 2E1 [5] | 3.55 | 0.8 | 0.4 | Flat | 6 |
| SS Mini 2S2, SS Mini 2E2 [5] | 3.55 | 0.8 | 0.6 | Flat | 2 |
| SS-1 [49] | 20.32 | 1.54 | 0.76 | Flat | 6 |
| SS-2 [49] | 12.7 | 1.13 | 0.25 | Flat | 6 |
| SS-3 [49] | 7.62 | 1.54 | 0.76 | Flat | 6 |
| SS-J, SS-J3 [5], [49] | 5 | 1.2 | 0.75 | Flat | 11 |
| SS-Mini [49] | 2.3 | 0.4 | 0.25 | Flat | 6 |
| Rund et al. (2015) [33] | 15 | 5 | 0.5 | Flat | 12 |
| Džugan et al. (2015) [59] | 3 | 1.5 | 0.5 | Flat | 19 |
| Byun et al. (1998) [56] | 10 | 3 | [0.12:2] | Flat | 11 |
| Alsabbagh et al. (2013) [58] | 2 | 0.2 | 1 | Flat | 2 |
| Džugan et al. (2015) [59] | 10 | [4.84, 30, 7.99, 9.88, 5.01] | [4.84, 30, 7.99, 9.88, 5.01] | Round | 7 |
| Van Zyl et al. (2016) [60] | 10 | 2 | 1 | Flat | 8 |
| A80 [13] | 80 | 20 | [1.2, 1.3, 1.6] | Flat | 5 |
| A50 [13] | 50 | 12.5 | [1.2, 1.3, 1.6] | Flat | 5 |
| ASTM25 [13] | 25 | 6 | [1.2, 1.3, 1.6] | Flat | 5 |
| Mini1 [13] | 10 | 3 | [1.2, 1.3, 1.6] | Flat | 5 |
| Mini2 [13] | 5 | 2 | [1.2, 1.3, 1.6] | Flat | 5 |
| Type I [61] | 30 | 6 | 6 | Round | 13 |
| Type II [61] | 9.5 | 3 | 1 | Flat | 6 |
| Type III [61] | 3 | 1 | 0.3 | Flat | 13 |
| Balakrishnan et al. (2014) [62] | [8.6 22] | [1.3 6] | [2.4 3.6, 4.3] | Flat | 22 |
| Konopík et al. (2018) [63] | [4 50] | [1 1.5] | [0.2 0.5 1.5] | Flat | 12 |
| Lanjewar et al. (2019) [64] | 3 | 1 | 1.3 | Flat | 3 |
| Dymáček et al. (2018) [54] | 4 | 2 | 1 | Flat | 2 |
| Dymáček et al. (2018) [54], Do Kweon (2021) [71], Mishra et al. (2021) [72] | 25 | 5 | 5 | Round | 28 |
| Sikan et al. (2023) [65] | 3 | 1 | 0.5 | Flat | 3 |
| Klueh (1985) [50] | [7.62 12.7 20.32] | [1.02 1.52] | [0.25 0.76] | Flat | 54 |
| Klueh (1985) [50] | 18.3 | 2.03 | 2.03 | Round | 18 |
| Igata et al. (1986) [66] | [15 30] | 4 | [0.0032 0.1 0.185 0.2 0.35] | Flat | 39 |
| Roach et al. (2020) [68] | [1.6, 2.4, 4, 10, 25] | [0.4, 0.6, 1.0, 2.5, 6.25] | [0.4, 0.6, 1.0, 2.5, 6.25] | Round | 5 |
| Yin et al. (2023) [70] | [6, 10, 15] | [3.2, 4, 6.4] | [0.5, 1, 2] | Flat | 22 |
| Do Kweon (2021) [71] | 25 | 6 | 6 | Round | 2 |
| Miyahara et al. (1985) [69] | 15 | 4 | [0.02, 0.1, 0.15, 0.2, 0.3, 0.35, 0.5] | Flat | 238 |
| ASTM 1:1 [57] | 10 | 10 | 0.64 | Flat | 1 |
| ASTM 3:2A [57] | 15 | 10 | 0.64 | Flat | 1 |
| ASTM 3:2B [57] | 15 | 10 | 0.64 | Flat | 1 |
| ASTM 4:1 [57] | 40 | 10 | 0.64 | Flat | 1 |
| ASTM A [49] | 28 | 6 | 6 | Flat | 6 |
| Kohno et al. (2000) [36] | 5 | 1.2 | [0.06, 1.03] | Flat | 421 |
| Seo et al. (2022) [85] | [5, 20, 38] | [3, 5] | [3, 5] | Round | 17 |

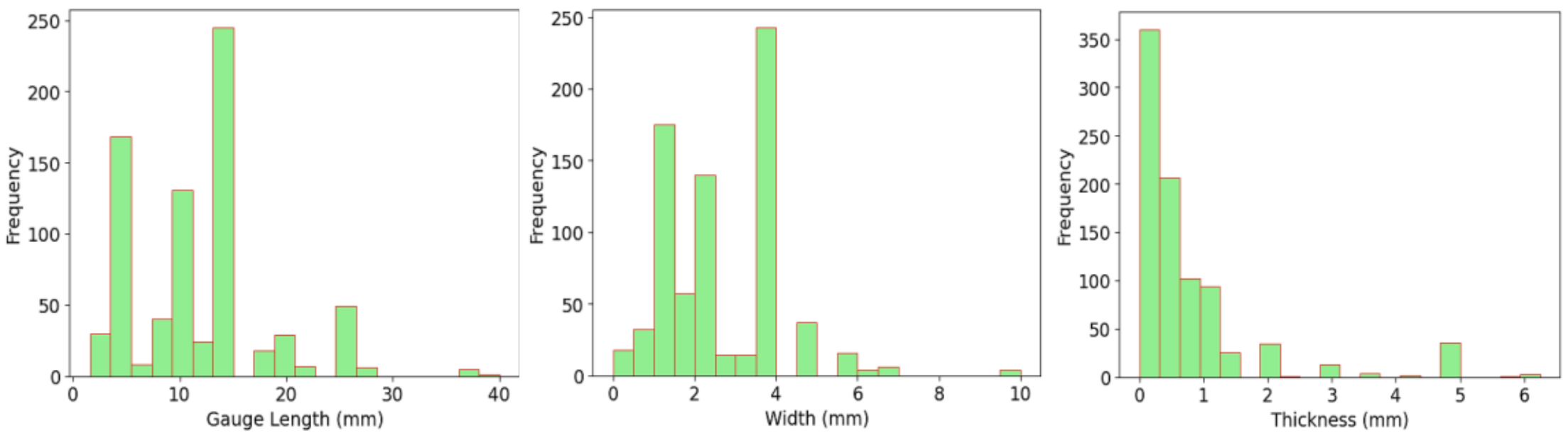

(a) (b) (c)

**Fig. 5.** Histograms for the dimensions of sub-sized specimens: (a) gauge length, (b) width or diameter, (c) thickness or diameter.

### Tensile Properties

The collected data for the tensile properties encompass information for yield strength, ultimate tensile strength, uniform elongation, and total elongation. The units for yield strength and ultimate tensile strength are MPa (Mega Pascals), and percent (%) for elongation. The distributions of the tensile properties for the sub-sized specimens in the dataset are shown in the histograms in Figure 6.

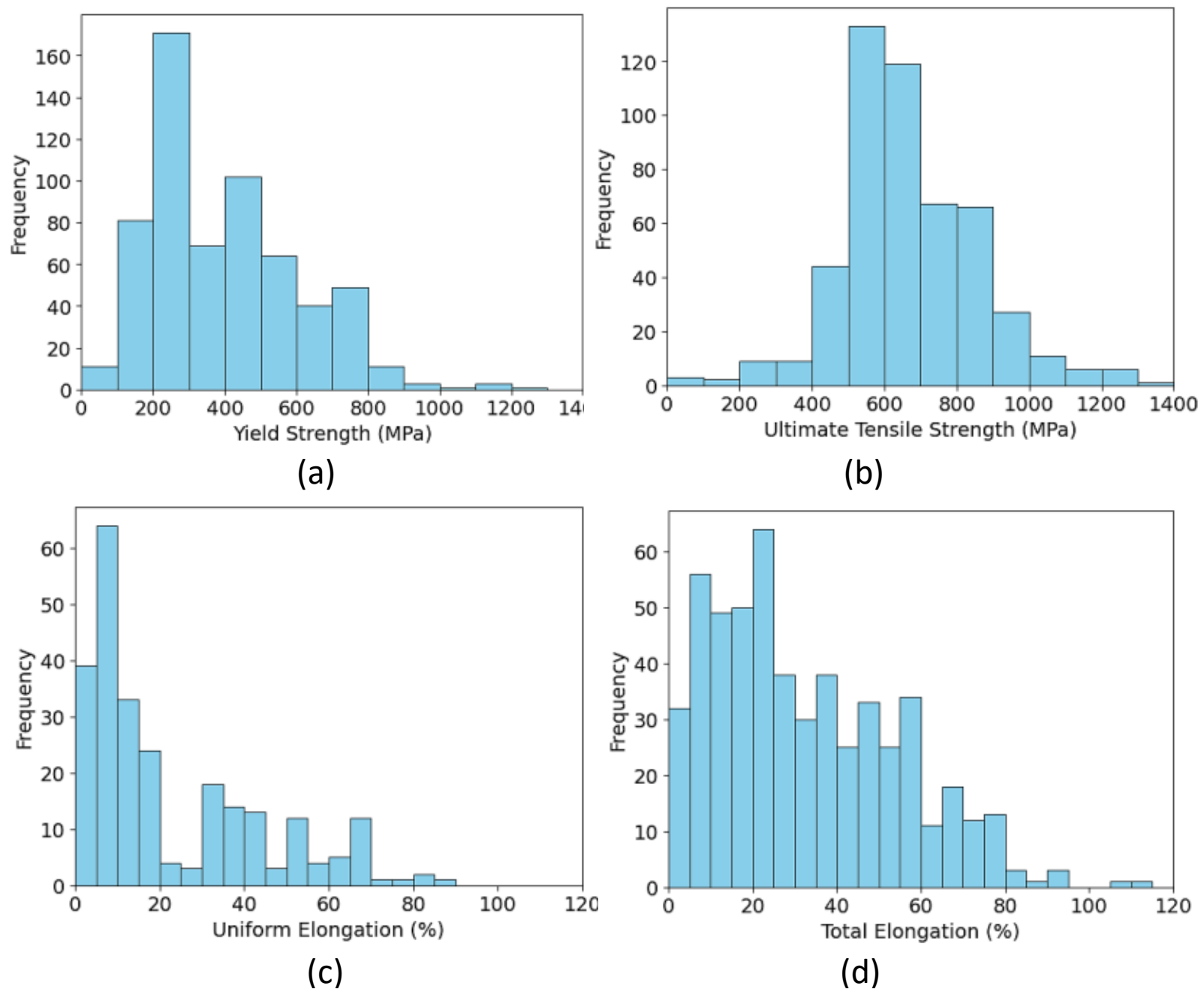

**Fig. 6**. Histograms of tensile test properties: (a) yield strength, (b) ultimate tensile strength, (c) uniform elongation, and (d) total elongation.

### Manufacturing Treatment and Irradiation Conditions

Histograms of the distribution of manufacturing treatment of the specimens and irradiation conditions related to the irradiation dose and temperature are presented in Figure 7.

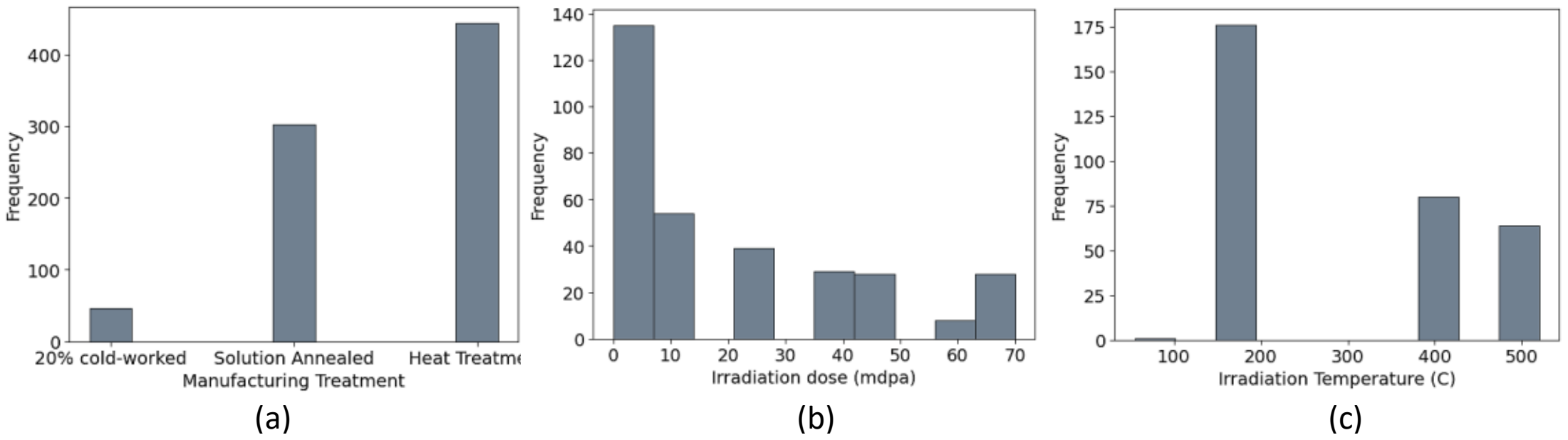

**Fig. 7**. Histograms of (a) manufacturing treatment (b) irradiation dose, and (c) irradiation temperature.

## Technical Validation

To ensure data validity, materials science experts conducted a thorough review of all retrieved data from the literature. This step involved ensuring the accuracy of the material type and use of standardized material names and verifying the chemical composition of the materials in the dataset. The experts validated the information for the manufacturing processes and treatment methods, testing conditions, specimen size, and tensile test properties. Similarly, the conversion of extracted information to consistent units was validated. Furthermore, our team of experts performed statistical analyses and applied data visualization techniques to identify and address data outliers, ensuring the reliability of the dataset.

We analyzed the data to validate the MTT recommendations for sub-sized specimens to ensure that the material bulk properties are maintained and that analytical data techniques can be applied to account for the specimen size effect. Most sub-sized specimens selected for this dataset have thickness of 0.2 mm or greater (to ensure containing at least 10 grains) [7], however we elected to include data from several thinner specimens with thicknesses of: 0.02, 0.1, 0.06, and 0.15 mm. It is important to note that these thickness values are part of a study where the authors varied the thickness of the specimen to evaluate the impact on tensile properties [61], and are not actual specimen designs used in practice. We didn't discard these data points although they violate the recommendations for minimum thickness, because we believe that keeping the data can be helpful for studies of the specimen size effect. As stated earlier, recommendations for the ratio of width-to-thickness $W/T$ are to not be lower than a critical value of around 5. The histogram of the width-to-thickness ratio is shown in Figure 8(a). Similarly, for the ratio of the gauge length to the square root of the cross-sectional area $L_0/\sqrt{A}$, the recommended critical value is 5.65. The histogram of the $L_0/\sqrt{A}$ ratio is plotted in Figure 8(b), showing that 79% of the records have the recommended aspect ratio. The recommended values for the length-to-width ratio $L_0/W$ for sub-sized specimens are typically in the range from 3 to 6, with most specimen designs having a value of approximately 4 [7]. Based on the histogram of the $L_0/W$ ratio shown in Figure 8(c), 87% of the data records fall within that range. Some specimen geometries (such as SS-1 and SS-2) have a length-to-width ratio over 10, whereas some other designs (such as ASTM 1:1 and ASTM 3:2) have smaller aspect ratios of 2 and below. We retained these data records, because they represent valid sub-sized specimen designs, and they can benefit studies of the specimen size effect.

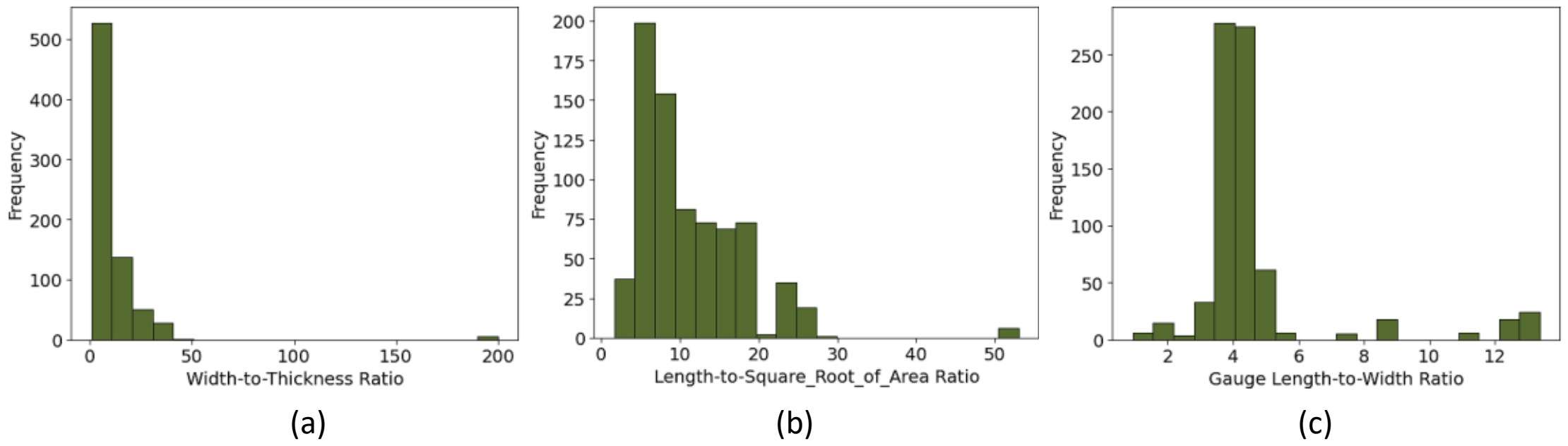

**Fig. 8**. Histograms of aspect ratios for the sub-sized specimens in the dataset: (a) width-to-thickness ratio, (b) length-to-square-root-of-area ratio, (c) gauge length-to-width ratio.

## Usage Notes

The primary usage of the dataset is to investigate the specimen size effect on the tensile properties of nuclear structural materials. Specifically, the objective is to enable researchers and engineers to apply statistical analysis methods and ML models for studying the size effect on specimens with different sizes and geometries. The dataset is conveniently structured for implementing conventional ML models and deep learning-based models for predicting the tensile properties of sub-sized specimens, including yield strength, ultimate tensile strength, uniform elongation, and total elongation. Loading the dataset and performing exploratory data analysis can easily be done by using standard python libraries, such as pandas, matplotlib, and scikit-learn, as demonstrated in the example code (link provided in the next section). A related purpose of the dataset is to establish statistical correlations between the tensile properties of sub-sized and standard-sized specimens. Indeed, the next goal of our team is to develop a web API for the potential users of the dataset. The API will allow users to query the database by submitting information regarding specimen size, material type and composition, and tensile testing environment, and in response they will receive predictions of the tensile test results and properties along with uncertainty estimates for the predicted values.

Another usage of the dataset is for investigating the impact of various test factors on the tensile properties of sub-sized specimens. To this end, statistical analysis methods can be applied to the tensile test records to extract insights about the impact and significance of different input parameters on specific tensile properties. Alternatively, interpretable and explainable ML methods can be implemented for quantifying the contribution of different features, parameters, and conditions to the decision-making process of the models in predicting tensile properties.

Other potential use of the dataset includes design of sub-sized specimens for new materials, whereas by extrapolating the predicted tensile properties for different specimen geometries, the results can be facilitated for determining the optimal size and volume of sub-sized specimens. Additionally, the data can be used for experiment design in tensile tests by identifying the most informative testing conditions that lead to reduced testing cycles for new materials. A promising approach for this task includes the development of active learning ML methods for identifying testing conditions with high data informativeness and representativeness in order to guide the test matrix design for advanced materials.

Conclusively, the dataset can represent a valuable resource for the community, as we are not aware of any similar datasets in the published literature that provide open-source structured data for sub-sized specimens. Our team will continue to improve and expand the dataset by adding new tensile test records and additional mechanical properties for sub-sized specimens. Considering the importance of the specimen size effect on the properties of sub-sized specimens, this work can pave the way for the development of novel data-driven analysis methods and for accelerated qualification of advanced materials.

## Code Availability

Sample code in Python and Jupyter Notebooks for loading the dataset and analysis of the results are available at the following link: https://github.com/avakanski/Subsized-Specimens-Tensile-Properties/blob/main/Sample_Code.ipynb.

## Acknowledgments

This work was supported through the INL Laboratory Directed Research& Development (LDRD) Program under DOE Idaho Operations Office Contract DE-AC07-05ID14517 (project tracking number 24A1081-149). Accordingly, the publisher, by accepting the article for publication, acknowledges that the U.S. Government retains a nonexclusive, paid-up, irrevocable, worldwide license to publish or reproduce the published form of this manuscript or allow others to do so, for U.S. Government purposes.